\documentstyle[mprocl,epsf]{article}

\def\be{\begin{equation}}
\def\ee{\end{equation}}
\def\bea{\begin{eqnarray}}
\def\eea{\end{eqnarray}}

\def\be{\begin{equation}}
\def\en{\end{equation}}
\def\bear{\begin{eqnarray}}
\def\enar{\end{eqnarray}}
\def\beas{\begin{eqnarray*}}
\def\enas{\end{eqnarray*}}
\def\bera{ \setcounter{enumi}{\value{equation}} 
           \addtocounter{enumi}{1}
           \setcounter{equation}{0}
           \renewcommand{\theequation}{\theenumi\alph{equation}}
           \begin{eqnarray} }
\def\enra{ \end{eqnarray}
           \setcounter{equation}{\value{enumi}}
           \renewcommand{\theequation}{\arabic{equation}}  }

\def\nonn{\nonumber \\ &&}

\def\mova{\left( M / a\right)}

\def\been{\begin{enumerate}}
\def\enen{\end{enumerate}}
\def\beit{\begin{itemize}}
\def\enit{\end{itemize}}

\def\bece{\begin{center}}
\def\ence{\end{center}}
\def\bert{\begin{flushright}}
\def\enrt{\end{flushright}}

\begin{document}
\renewcommand{\thefootnote}{\fnsymbol{footnote}}
\setcounter{footnote}{1}



\title{Newtonian and Post-Newtonian Binary Neutron Star Mergers
\footnote{The Proceedings of the 8th Marcel Grossmann Meeting,
Jerusalem, June 1997 (World Scientific Press); gr-qc/9710073 }}

\author{
 Hisa-aki Shinkai$^{1}$,
 Wai-Mo Suen$^{1}$,
 F. Douglas Swesty$^{2,3}$
 Malcolm Tobias$^{1}$, \\
 Edward Y. M. Wang$^{1,2,4}$,
 Clifford M. Will$^{1}$
}
\address{
 $^{1}${Dept. of Physics, Washington Univ., St. Louis, 
        MO63130-4899, USA}\\
 $^{2}${Lab. for Computational Astrophysics,
        Univ. of Illinois, Urbana-Champaign, IL , USA}\\
 $^{3}${Dept. of Astrophysics,
        Univ. of Illinois, Urbana-Champaign, IL , USA}\\
 $^{4}${National Center for Supercomputing Applications,
         Urbana-Champaign, IL , USA}
}

\maketitle\abstracts{
 We present two of our efforts directed toward the numerical analysis of
 neutron star mergers, which are the most plausible sources for 
 gravitational wave detectors that should  begin operating in the near future.
 First we present Newtonian 3D simulations including radiation
 reaction (2.5PN) effects.  
 We  discuss the 
 gravitational wave signals and luminosity from the merger
 with/without radiation reaction effects.  
 Second we present
 the matching problem between post-Newtonian formulations and 
 general relativity in numerical treatments. 
 We prepare a spherical, static neutron star in a post-Newtonian 
 matched spacetime, and find that  discontinuities at the matching 
 surface become smoothed out 
 during
 fully relativistic evolution if we use a proper slicing condition.}



\renewcommand{\thefootnote}{\arabic{footnote}}
\setcounter{footnote}{0}

\section{Introduction}

 Mergers of neutron stars in binary systems
 are considered excellent astrophysical
 laboratories for  gravitational wave astronomy, nuclear
 astrophysics and relativistic astrophysics. 
 The gravitational waves emitted from these mergers are expected 
 to be observed by
 gravitational wave detectors coming on-line in the next decade, such
 as LIGO, VIRGO, GEO and TAMA.  
 Detailed predictions of waveforms are desired and need to be
 produced analytically and numerically for the extraction of 
physical information.

  Post-Newtonian (PN) calculations \cite{will94} 
  can accurately describe the evolution of the system when the 
  separation between the two stars in the binary system 
  is much larger than the stellar radii. 
  The final phase of the neutron star mergers (NSM)
  requires fully general relativistic (GR)
  numerical simulations.
In this paper, we  
 present two of our efforts 
at building a bridge between these two approaches.

\section{Newtonian simulation of NSM including radiation reaction}

 First we present  Newtonian 3-dimensional simulations including radiation
 reaction (2.5PN) effects \cite{wang-swesty}. 
 We note that the 
 Kyoto \cite{shibata92} and Max-Planck \cite{ruffert} groups 
 have also presented results of their NSM simulations using
 Newtonian equations plus 2.5PN corrections. 
 We use the same set of equations,
which are a 
reduction of the PN-hydro formulation of  Blanchet, Damour and Sch\"afer
\cite{BDS}.
We are studying 
various 
 initial conditions and  comparing the effects of different 
 equations of state.

 We evolve the Euler equations 
 using a modification of the ZEUS-2D algorithm,
 staggered grid structure, 
 second order van Leer monotonic interpolator, and Norman's
 consistent advection method \cite{stone-norman}.
  We have conducted convergence studies to delineate how spatial and
  temporal resolution affect the conservation of angular momentum and
  energy in these models. 
In order to get sufficient  angular momentum conservation,
we find that we must update the ``advection" terms before solving the Poisson
equations.  To solve  the  Poisson equations, we use a 
full multi-grid W-cycle algorithm.

Here, we show an example of the gravitational waveform emitted by the
coalescence of two equal mass stars. The equation of state is
polytropic $P=K \rho^2$  with initial mass $1.4 M_\odot$ and 
radii $ R_\ast=9.56{\rm Km}$.
The stars are separeted initially by  $2.9 R_\ast$, are nonrotating, 
and have an orbital velocity taken to be the Kepler velocity with 
infalling radial velocity determined by radiation reaction 
This simulation was made with $129^3$ grid zones ($\Delta x=0.893$Km) 
and required  $160$ hrs  of CPU time 
by Origin 2000 with an average time step 
$1.43 \times 10^{-3} $ms for a total physical
time of 5.40 ms.  

In Fig.1, we show the gravitational waveform $h_\times$ and luminosity
with and without radiation reaction term. We see that the merger
occurs earlier when we include radiation reaction terms. This 
may be explained by the loss of angular momentum from 
the system induced by radiation reaction. 
Detail reports are now in preparation \cite{wang-swesty}.

\setlength{\unitlength}{1.0in}
\begin{picture}(4.8,2.3)
\put(0.0,0.25){\epsfxsize=2.0in \epsfysize=2.0in \epsffile{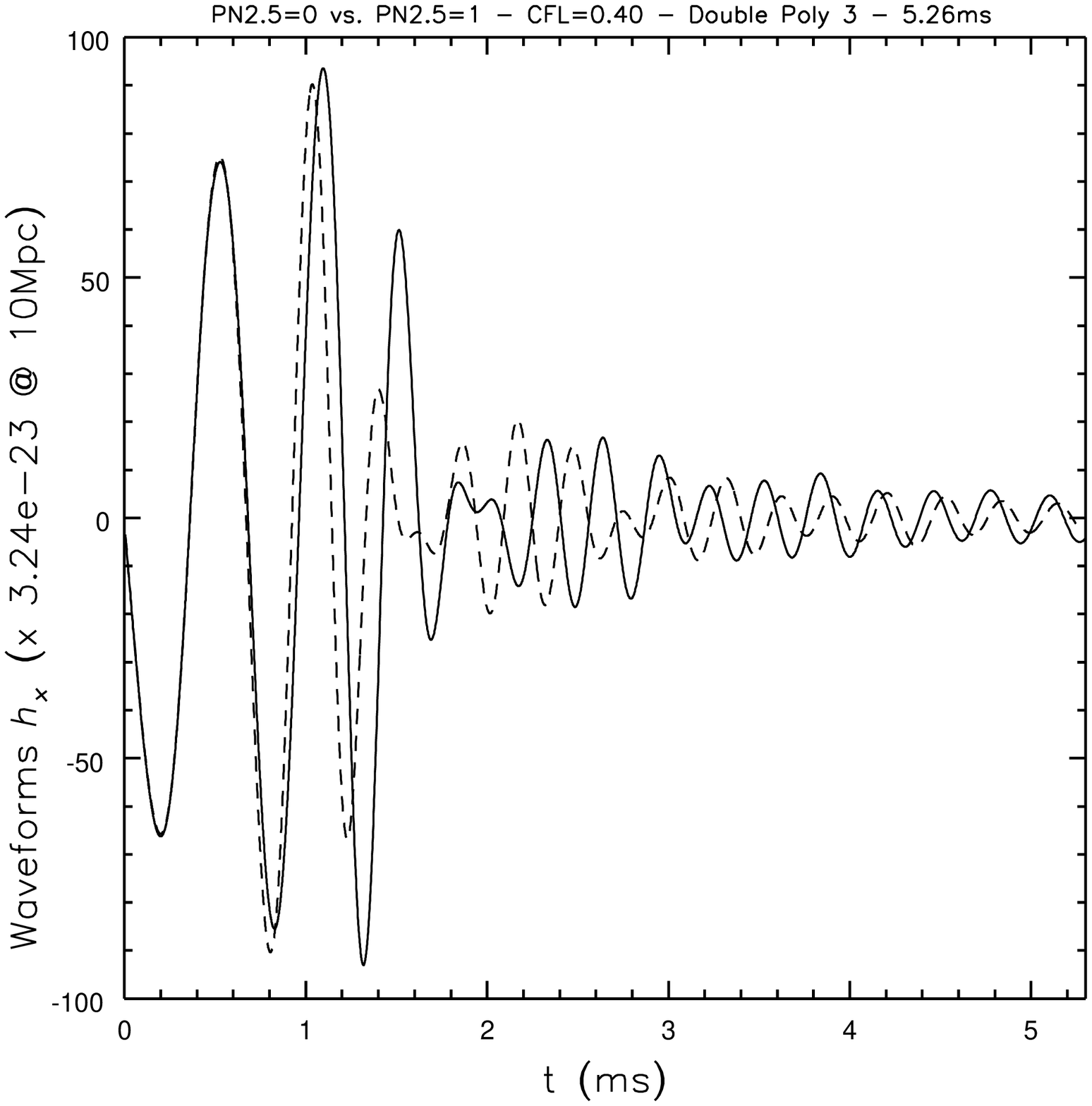} }
\put(2.2,0.25){\epsfxsize=2.0in \epsfysize=2.0in \epsffile{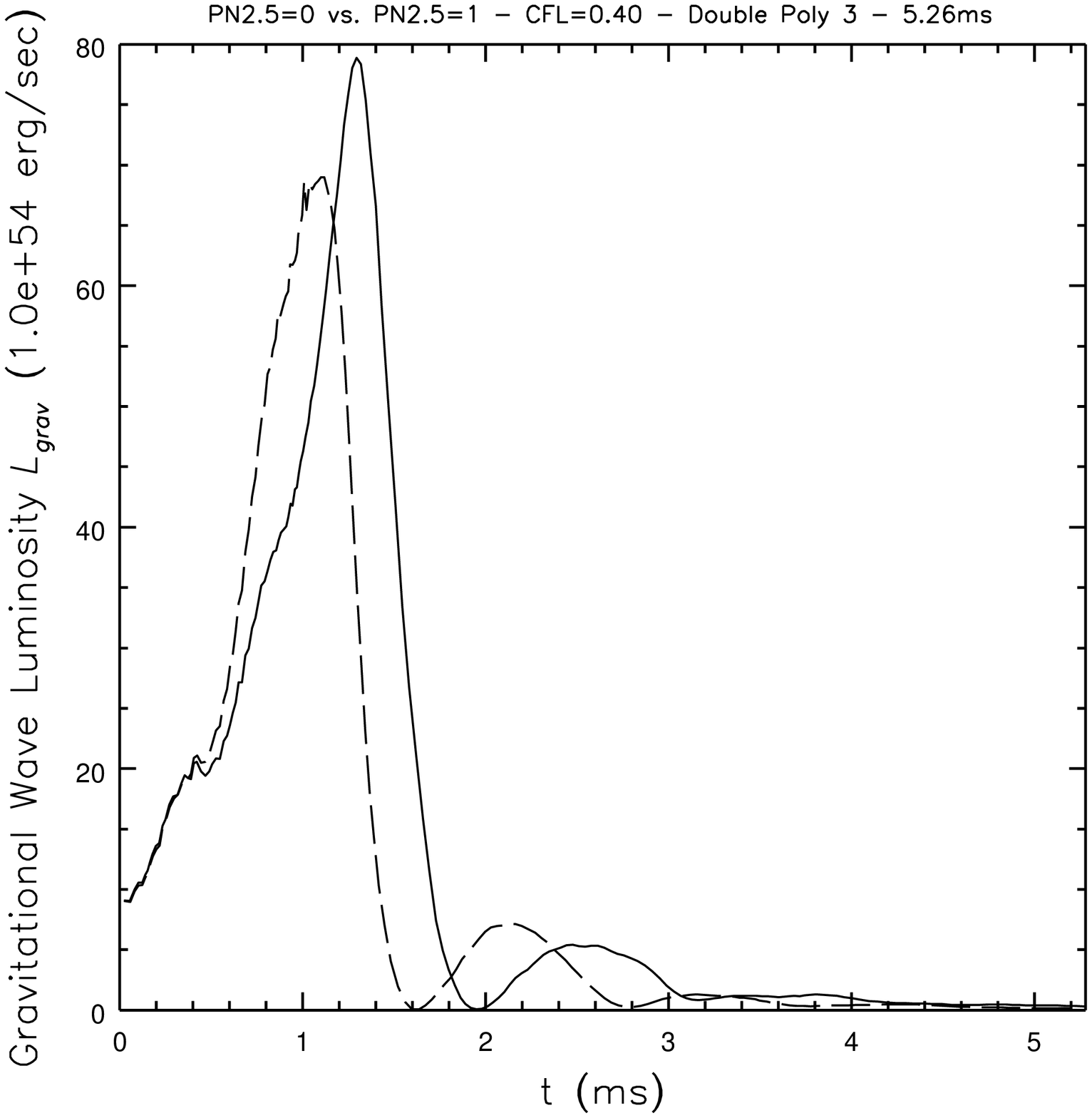} }
\put(0.2,0.0){\parbox[t]{4.0in}{Fig. 1:
Waveform $h_\times$ and luminocity. Solid line is Newtonian and 
dashed line is with radiation reaction.
}}
\end{picture}

~

\section{GR evolution of a neutron star 
with PN matched initial data}
 
Our second effort in linking 
PN calculations to full GR numerical simulations involves construction
of PN initial data for fully relativistic numerical treatment.

As a first step, we constructed 
a single spherically
symmetric neutron star by solving the Tolman-Oppenheimer-Volkoff equations;
this solution is matched to a 1PN vacuum metric at 
some surface outside the star. 
We then 
evolved this initial data using a fully relativistic  spherical
code, and investigated  what happens at the matching surface.  

In the central region of the star, we set our fully relativistic
metric to
\begin{equation}
ds^2 = - e^{2\Phi(R)} dt^2 + e^{2\Lambda(R)} dr^2 + R^2(r)\left( d\theta^2
+ \sin^2 \theta d\varphi^2 \right),
\end{equation}
where $r=r(R)$. Under the harmonic gauge condition
$ {d \over dr}
\left[ e^{\Phi - \Lambda} \rho^2   \right]  = 2r e^{\Phi +\Lambda},
$
we solve the hydrostatic equilibrium (TOV) equation and field
equations assuming a polytropic equation of state. We define a
``mass'' $M$,
given by $ M=4 \pi \int_0^{R_{\ast}} \rho(R) R^2 dR$, where $R_\ast$
is the radius of the star. We set the 
exterior metric of the neutron star by 
$ \Phi(R)={1\over 2} ln(1-{2M\over R})+C, $ and  
 $ e^{2\Lambda}=(1-{2M\over R})^{-1}$,
 where $R=r+M$ and 
  the constant term $C$ is to be fixed by matching later.
The outer metric (PN region; $r \ge a \ge R_\ast$, where $a$ is the matching
radius) is given by 
\bear
ds^2&=&-(1-2U+2U^2+\beta U^3)dt^2
+[1+2U+2U^2+{2\over 3}\alpha U^3]dr^2 \nonn
+[1+2U+U^2-{1\over 3}\alpha U^3]r^2(d\theta^2 +\sin\theta d\phi^2) \,,
\enar
where  $ U \equiv{\tilde{m} / r}$, with $\tilde m$ the Kepler-measured
mass of the star, $\beta$ the coefficient of a 2PN term, and $\alpha$
denoting a residual gauge freedom within harmonic gauge.  Matching
inner and outer metrics at $r=a$ in the standard way yields
\bera
\tilde{m}&=&M [ 1+ \mova^2+4\mova^4+O(\mova^5)] \,,\\
\alpha&=&6[1+\mova+ \mova^2+2\mova^3+O(\mova^4)] \,,\\
\beta&=&-{4 \over 3}[1+\mova-2\mova^2+O(\mova^3)] \,,\\
C&=&-{2 \over 3}\mova^3-{2 \over 3}\mova^4-{37 \over 15}
\mova^5+O(\mova^6)] \,.
\enra
This initial data contains  a discontinuity in the second derivative of
the metric at the matching surface $r=a$.  After fully relativistic
evolution, we find this matching discontinuity to be smoothed
out if we use the maximal slicing condition or ``K-driver" slicing
condition \cite{Kdriver}, both of which are given by solving an elliptic
equation for the lapse function.  
 ``K-driver" slicing  smoothed out the matching 
discontinuities faster than the maximal slicing case. 
Other slicing conditions such as static lapse
(harmonic gauge) and the algebraic slicing condition $\alpha=1+\log
\gamma$ 
are shown to lead to high frequency noise 
caused by the discontinuity at the matching surface.

This study suggests that the matching surface is not necessarily fatal,
provided suitable slicing conditions are used. 
 Such matching surfaces
are expected to exist in constructing initial data using the PN formulation
for full GR evolution. Results from simulations using such a PN initial
data will be reported in the near future.


We thank Mike Pati and Nils Andersson for discussions. 
This work was supported in part by NSF PHY96-00507 and PHY 96-00049,
and by  NASA
ESS/HPCC CAN NCCS5-153.


\end{document}